\documentclass[12pt,a4paper]{iopart}

\usepackage{amssymb}
\usepackage{graphicx}
\usepackage{color}
\usepackage{hyperref}

\usepackage[numbers,sort&compress]{natbib}

\usepackage[utf8]{inputenc}

\newcommand{\mytitle}{Characterization of attosecond pulses in the soft x-ray regime}

\definecolor{MyDarkGreen}{rgb}{0,0.6,0}
\definecolor{MyDarkBlue}{rgb}{0,0,0.8}
\definecolor{MyDarkRed}{rgb}{0.6,0,0.3}
\hypersetup{breaklinks=true, colorlinks=true,plainpages=true, linktocpage=true, linkcolor=MyDarkBlue, citecolor=MyDarkGreen, urlcolor=MyDarkRed, pdfborder={0 0 0},%
pdfauthor={Stefan Pabst},%
pdfsubject={Research article},
pdftitle={\mytitle}%
}

\newcommand{\ket}[1]{\left|#1\right>}
\newcommand{\bra}[1]{\left<#1\right|}

\newcommand{\braket}[2]{\left<#1|#2\right>}

\newcommand{\eqref}[1]{(\ref{#1})}

\begin{document}

\title{\mytitle}

\author{Stefan Pabst$^{1,2,*}$ and Jan Marcus Dahlstr\"om$^{3}$}

\address{$^1$ ITAMP, Harvard-Smithsonian Center for Astrophysics, 60 Garden Street, Cambridge, Massachusetts 02138, USA}
\address{$^2$ Physics Department, Harvard University, 17 Oxford Street, Cambridge, Massachusetts 02138, USA}

\address{$^3$ Department of Physics, Stockholm University, AlbaNova University Center, SE-106 91 Stockholm, Sweden}

\date{\today}


\begin{abstract}
Attosecond x-ray pulses offer unprecedented opportunities for probing and triggering new types of ultrafast motion.
At the same time, pulse characterization of x-rays presents new challenges that do not exist in the UV regime. 
Inner-shell ionization is the dominant ionization mechanism for x-rays and it is followed by secondary processes like fluorescence, Auger decay, and shake-up.
In general, we find that inner-shell ionization and secondary processes can create additional delay-dependent modulations that will affect pulse reconstruction schemes.  
Our recently proposed pulse characterization method [Pabst and Dahlstr\"om, PRA {\bf 94}, 013411 (2016)], where a bound electron wavepacket is sequentially photoionized by the attosecond pulse, can be adapted to mitigate the impact of these effects, thus opening up an avenue for reliable pulse reconstruction in the x-ray regime. 
\end{abstract}

\pacs{32.80.-t,42.65.Re,31.15.vj,32.80.Ee}
\maketitle


\section{Introduction}
\label{sec:intro}

The first attosecond pulses were created and probed in 2001~\cite{HeKr-Nature-2001,PaAg-Science-2001}. 
This signaled the birth of a new sub-field of physics known as attosecond physics~\cite{KrIv-RMP-2009}. 
Today, the shortest reported attosecond pulses have a duration of 67~as~\cite{ZhZh-OptLett-2012}, 
which is much shorter than the oscillation period of optical light ($\sim$1~fs). 
To reach these short pulse duractions it is, therefore, necessary to rely coherent UV or x-ray fields. 
In the last years, coherent attosecond pulses in the soft x-ray regime ($\omega\gtrsim 300$~eV) 
with bandwidths of more than 100~eV have been produced~\cite{PoCh-Science-2012,IsIt-NatComm-2014,SiBi-NatComm-2015,TeBi-NatComm-2016}.

On the one hand, these pulses begin a new chapter of how we can probe inner-shell electronic and nuclear motion. 
Besides the unprecedented temporal resolution, x-rays offer the ability to study the electronic environment around 
a specific atomic site within the material~\cite{Si-RMP-XrayReview-1982}, and brings core-hole spectroscopy 
into the attosecond regime~\cite{PfSp-RPP-2006}. 
Attosecond x-ray pulses may find many applications in studies ranging from multi-orbital electronic dynamics~\cite{SaTi-NatPhys-2017}
non-Born-Oppenheimer dynamics~\cite{LiVe-PRL-2015,PeWo-Science-2017}, charge-transfer processes in photochemical reactions~\cite{CrWu-Nature-2005} 
to structural~\cite{CoTe-Science-2003} as well as insulator-metal~\cite{GlYo-PRL-2003,CaSc-PRL-2005} phase transitions in condensed matter systems.

On the other hand, the characterization of broad attosecond x-ray pulses faces new challenges that do not exist in the UV regime.
X-rays predominantly ionize inner-shell electrons creating a highly excited ion.
The extremely large bandwidths of these attosecond x-ray pulses exceed the energy gap between many atomic shells 
making it energetically impossible to distinguish from which shell the photoelectron was ionized. 
While this indistinguishability creates problems for the pulse characterization, as spectral components 
from very different spectral regions of the pulse contribute to the same final photoelectron energy, it also  
opens up the possibility to prepare coherent hole wave packets with large energy spacing and dynamics the attosecond time scale~\cite{PaSa-PRL-2011}. 

The spectral phase of an isolated attosecond pulse is commonly determined by 
the FROG-CRAB method\footnote{frequency-resolved optical gating-complete reconstruction of attosecond bursts}~\cite{MaQu-PRA-2005}, 
where the photoelectron continuum is dressed with an IR pulse that acts as a phase gate 
to retrieve the UV pulse shape.  
PROOF\footnote{phase retrieval by omega oscillation filtering}~\cite{ChGi-OptEx-2010-PROOFMethod} 
is an alternative method where a weaker IR pulse is used to create interference in the photoelectron spectrum between two distinct ionization pathways 
(UV-only and UV+IR ionization) that beat with the IR frequency 
as a function of the delay between UV and IR fields. 
Finally, RABBIT\footnote{Reconstruction of attosecond beating  by interference of two-photon transition (RABBIT)}~\cite{PaAg-Science-2001} is a pulse characterization technique that also uses similar photoelectron interferometry, but it is designed specifically for periodic trains of attosecond pulses.  

For extremely broad x-ray pulses, where the ionization from different shells cannot be energetically distinguished, the total delay-dependent modulation becomes an incoherent average over the modulations in each sub-channel.
Furthermore, the inner-shell holes are not stable and can decay via Auger decay or fluorescence.
To make things worse---before the hole decays it can change the state of outer electrons (shake-up and shake-off).
It has been already shown that shake-up affects the attosecond time delay~\cite{PaFe-PRL-2012} 
and it will consequently also affect the pulse reconstruction procedures.   
Already, experimental groups have troubles characterizing their attosecond x-ray pulses 
with the above mentioned pulse characterization methods~\cite{discuss-biegert}.

Recently, we have proposed a different pulse characterization method, 
which we will refer to as Pulse Analysis by Delayed Absorption (PADA), 
that is based on ionization of bound wavepackets \cite{PaDa-PRA-2016}. 
The different binding energies of the states involved in the wavepacket enable for spectral shearing interferometry of the photoelectron.  
The main differences between this method and those mentioned above is that: 
(i) the pump and probe steps are sequential; 
(ii) the intermediate states are bound; and 
(iii) the photoelectron is measured over all angles. 
These key distinctions allows for elimiation of the dipole phase contributions making the PADA method exact, 
with no associated delay due to the measurement procedure, at least within a one electron model.  
Finally, we mention that the development of the PADA method was inspired by theoretical work on non-sequential 
stimulated hole transitions to induce spectral shearing of photoelectrons \cite{YoDa-PRA-2016}.  
For all reconstruction methods (including PADA) it is important that the control over the pump-probe delay is significantly better than the pulse duration of the attosecond pulse and both pulses are phase-locked.


In this work, we discuss the challenges that arises when characterizing x-ray attosecond pulses and how most of these new challenges can be prevented with our wavepacket approach by choosing an advantageous wavepacket.
In \sref{sec:theory.idea} we explain the main idea of using bound wavepackets to characterize the pulse.
The influence of inner-shell ionization is discussed in \sref{sec:core}.
The impact of secondary processes like fluorescence, Auger decay, and shake-up on the photoelectron spectrum is discussed in \sref{sec:secondary}.
%
Atomic units (a.u.) are used throughout unless otherwise indicated.

\section{Basic Idea}
\label{sec:theory.idea}

The main challenge in characterizing a pulse, $\tilde{\cal E}(t)$, 
is the determination of the spectral phase, $\phi(\omega)$. 
The spectral phase contains information about the superposition of the different frequencies of the pulse,    
\begin{eqnarray}
  \label{eq:pulse}
  \tilde{\cal E}(t)
  &=
  \frac{1}{2\pi} \int_{-\infty}^\infty \! d\omega \ |{\cal E}(\omega)|\ e^{-i\omega\,t + i\phi(\omega)}
  ,
\end{eqnarray}
where $\tilde{\cal E}(t)$ is the pulse in the time domain and ${\cal E}(\omega)$ is the pulse in the spectral domain.  
Ionization is unavoidable when test pulses in the UV or x-ray regime interact with matter. 
Current attosecond pulse characterization techniques make use of this fact and determine the spectral phase via 
laser-assisted photoelectron spectra \cite{MaQu-PRA-2005,ChGi-OptEx-2010-PROOFMethod,PaAg-Science-2001}. 
In short, the laser light creates two or more possible ionization pathways 
(spectral interferometer arms) that depend on different spectral parts of the test pulse, 
thus encoding the spectral phase difference onto the photoelectron distribution.    

\begin{figure}[ht]
  \centering
  \includegraphics[width=\linewidth]{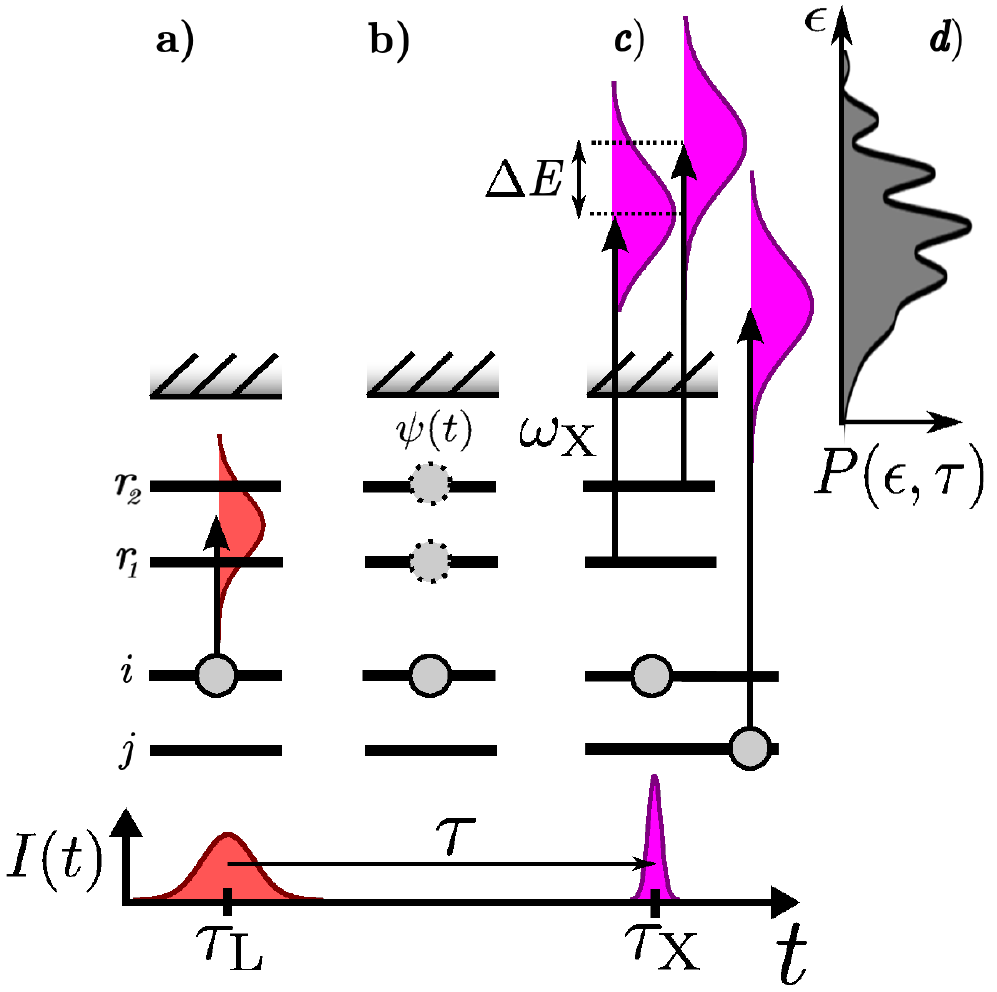}
  \caption{(color online) Sketch of the method to characterize broad pulses using a coherent electronic wavepacket.
  (a) Preparation of the wavepacket $\psi$.
  (b) Field-free propagation of the wavepacket for the duration $\tau$.
  (c) Ionization of the outer-shell wavepacket or of an inner-shell electron by the attosecond x-ray pulse.
  d) The photoelectron spectrum contains the interference due to the wavepacket and contributions from inner-shell ionization.
  }
  \label{fig:idea}
\end{figure}

\begin{figure}[ht]
  \centering
  \includegraphics[width=\linewidth]{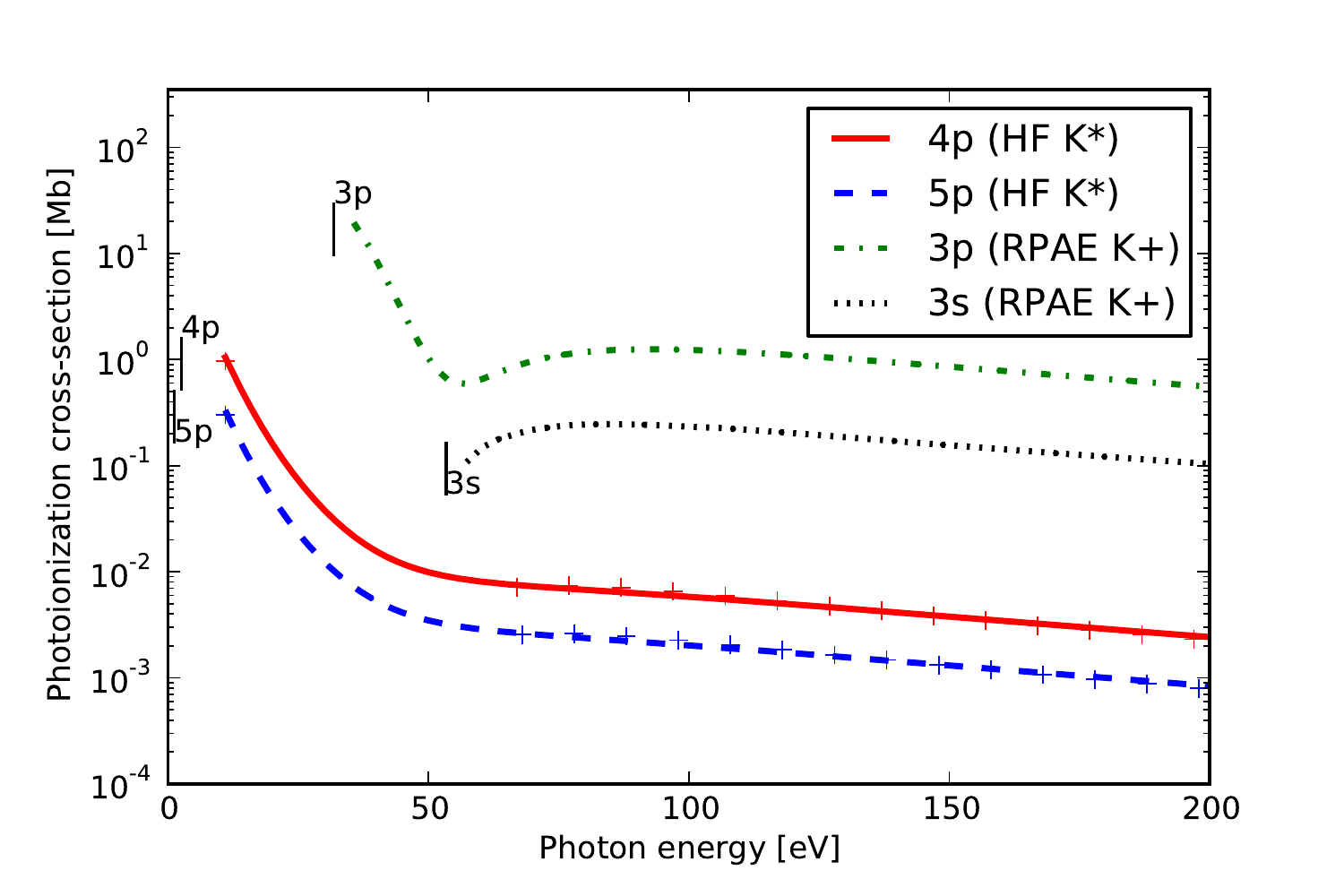}
  \caption{(color online) 
Partial photoionization cross-sections (PPCS) for excited potassium (K$^*$) and the potassium ion core (K$^+$).
  }
  \label{fig:K_CS}
\end{figure}

Recently, we have proposed the PADA method to characterize attosecond pulses~\cite{PaDa-PRA-2016}. 
The main idea is illustrated in Fig.~\ref{fig:idea}. 
Each excited bound state has a different ionization potential, $\Delta E=E_a-E_b$ and, consequently, 
different photon energies are needed to reach the same final photoelectron energy.  
In Fig.~\ref{fig:idea}c) it is shown that inner-shell electrons 
can be ionized to the same kinetic energy as the ionized outer electrons, 
provided that the x-ray pulses have an extremely broad energy width.  
The impact of this inner-shell ionization process will be the focus of this study. 
Since we did not consider explicitly inner-shell photoionization process in our earlier work on excited potassium~\cite{PaDa-PRA-2016}, 
we present in Fig.~\ref{fig:K_CS} the associated partial photoionization cross section (PPCS). 
The PPCS for the excited electron states $4p$ and $5p$ of potassium are computed by the Hartree-Fock (HF) method, 
while the PPCS of the potassium core K$^+$, from initially occupied orbitals $3p$ and $3s$, are computed 
by the random phase-approximiation with exchange (RPAE). 
The onset of ionization from $3p$ leads to slower photoelectrons that dominate in numbers by two or three orders of magnitude 
over the faster photoelectrons from the bound excited states. Indeed, if the UV/x-ray pulse is as broad as the $3p$ binding energy, 
then one should expect that the contrast of the excited bound ionization signal will be poor relative to the total amount of ionization. 
We have tested the validity of the Hartree-Fock calculation by adding the RPAE coupling of the inner-shell process to the photoionization 
of the Rydberg electron (indicated by $+$ signs in Fig.~\ref{fig:K_CS}) and find excellent agreement for photon energies not close to the opening 
of the inner shells in agreement with our earlier work \cite{PaDa-PRA-2016}.                
%

%
%
%

In RABBIT, PROOF and FROG-CRAB, the IR field affects all photoelectrons regardless of the ionic state, 
and, therefore, the spectral interferometer mechanism will affect every ionization channel. 
Interestingly, this is not the case in the PADA approach. 
%
%
To demonstrate the influence of inner-shell ionization for the PADA method,  
a two-level wavepacket between two eigenstates $\ket{E_a}$ and $\ket{E_b}$ is sufficient.
Note that a multi-level wavepacket is also possible and may be more favorable than a two-level wavepacket~\cite{PaDa-PRA-2016}.
The electron wavepacket, at the time of ionization $\tau$, is written as 
\begin{eqnarray}
  \label{eq:wvp}
  \ket{\Psi(\tau)}
  &=&
  g_{a} \ket{E_a} + g_{b} e^{i\Delta E \,\tau} \ket{E_b}
  ,
\end{eqnarray}
where $\Delta E=E_a-E_b$, and $g_{a/b}$ can be chosen to be real without loss of generality. 
In writing Eq.~\eqref{eq:wvp} we have assumed 100\% population transfer to the excited states, $|g_a|^2+|g_b|^2=1$.  
For noble gas atoms, weakly excited states (where an electron from the outer-most shell $i$ is excited into a Rydberg state $a$) 
are quasi stable and they do not decay on the time scale that is relevant in this work.   
Using second quantization, these states can be written as one-hole--one-particle ($1h$--$1p$) configurations 
$\ket{E_a}=\ket{\Phi^{a}_i}=\hat c^\dagger_{a}\, \hat c_i \ket{\Phi_0}$, 
where $\ket{\Phi_0}$ is the neutral ground state of the noble gas atom. 
The exact excitation energy can by appoximated by Koopman's theorem, $E_a=E^{a}_i = \epsilon_{a}-\epsilon_i + E_0$, 
where $\epsilon_p$ is the energy of the one-particle orbital $p$ and $E_0$ is the energy of $\ket{\Phi_0}$. 
The creation and annihilation operators of an electron in orbital $p$ is given by $\hat c^\dagger_p$ and $\hat c_p$, respectively.
The coefficients $a,b$ refer to unoccupied bound (Rydberg) orbitals, $c,d$ to continuum states, $p,q$ refer to Rydberg and continuum state orbitals, $i$ to the outer-most occupied orbital containing the primary hole, and $j$ to all other occupied orbitals.

While the exact details of the preparation of the bound wavepacket are not important, 
it is essential that the bound wave packet is coherent with the test pulse. 
In practice, the wave packet preparation process may involve non-linear interaction with ultrashort pulses 
to drive outer-valence electron population to the desired target excited bound states.   
In fact, it is also possible to use coherent hole wavepackets instead of Rydberg electron states.   
Previous studies have shown that coherent hole motion can be created via one-photon~\cite{PaSa-PRL-2011} and tunnel ionization~\cite{PaWo-PRA-2016}.

The final angle-integrated photoelectron spectrum as a function of photoelectron energy, $\epsilon$, is given by
\begin{eqnarray}
  \label{eq:pes.std}
  \hskip-5ex
  P(\epsilon,\tau)
  &=&
  \sum_{p=a,b}
  A_p(\epsilon) 
  +
  2B(\epsilon)  \cos\Theta(\epsilon,\tau)
  ,
\end{eqnarray}
where $A_p$ is the static contribution from state $\ket{E_p}$ and $B(\epsilon)$ is the strength of the interference between $\ket{E_a}$ and $\ket{E_b}$.
The phase of the interference is $\Theta(\epsilon,\tau)$, which is energy dependent and encodes the spectral phase information.
The coefficients read
\numparts
\begin{eqnarray}
  \label{eq:pes.std_coeff1}
  A_p(\epsilon)
  &=&
  g^2_p\ d^2_p(\epsilon)\  |{\cal E}(\epsilon-\epsilon_{p})|^2
  ,
  \\
  \label{eq:pes.std_coeff2}
  B(\epsilon)
  &=&
  g_{a}\, g_{b} \ d^2_{a,b}(\epsilon)\  |{\cal E}(\epsilon-\epsilon_{a})|\, |{\cal E}(\epsilon-\epsilon_{b})| 
  ,\\
  \label{eq:pes.std_coeff3}
  \Theta(\epsilon,\tau)
  &=&
  \textrm{arg}\Big( \Delta E \tau + [\phi(\epsilon-\epsilon_{b})-\phi(\epsilon-\epsilon_{a})] \Big)
  .
\end{eqnarray}
\endnumparts
%
To arrive at Eqs.\eqref{eq:pes.std}-\eqref{eq:pes.std_coeff3}, 
we assumed in Ref.~\cite{PaDa-PRA-2016} that the attosecond pulse ionizes 
only the Rydberg electron. Thus, the parent ion is in its ground state, 
$\ket{\Phi_i}=\hat c_i \ket{\Phi_0}$, and no inner-shell ionization has taken place.  
In general, inner-shell ionization can not be neglected unless  
(i) the photoelectron energies reached from the bound wave packet are well separated from the contributions from the inner shells or 
(ii) the photon energies of the test pulse are below the opening of any other photoionization channels. 
Clearly, any remaining population in the ground state after preparation of the bound wave packet  
will also contribute additional photoelectron background.     
In \sref{sec:core}, we discuss the case when the bandwidth of the attosecond x-ray pulses 
exceeds the energy separation between shells and inner-shell ionization cannot be neglected anymore.
The dipoles entering Eqs.\eqref{eq:pes.std_coeff1} and \eqref{eq:pes.std_coeff2} are,
\numparts
\label{eq:pes.std_dip2}
\begin{eqnarray}
  \label{eq:pes.std_dip2p}
  d^2_p(\epsilon)
  &=&
  \sum_{\sigma,l,m}   \bra{\epsilon\,l_{m,\sigma}} z \ket{p}^2
  ,
  \\
  \label{eq:pes.std_dip2pq}
  d^2_{p,q}(\epsilon)
  &=&
  \sum_{\sigma,l,m}   \bra{\epsilon\,l_{m,\sigma}} z \ket{p} \bra{q} z \ket{\epsilon\,l_{m_l,\sigma}}
  ,
\end{eqnarray}
\endnumparts
and are averaged over all degenerate final states---namely spin $\sigma$ and angular momentum $l$ and $m$ of the photoelectron.
Note that for spherical symmetric systems without correlation the one-particle dipoles, 
$\bra{\epsilon\,l_{m,\sigma}} z \ket{p}$ are real~\cite{PaDa-PRA-2016,Friedrich-AMO-book}.
In contrast, the photoelectron spectrum in a specific direction introduces a dipole phase dependence in the $\cos$-modulation in Eq.~\eqref{eq:pes.std}.
In Ref.~\cite{PaDa-PRA-2016} we found numerically that the PADA method was rather insensitive to correlation effects, such as a Fano resonance. 
This result may appear surprising at first glance, because a Fano resonance in the continuum is associated with a rapidly varying dipole phase 
in a spectrally narrow energy window that could potentially invalidate the PADA result.   
However, using Fano's theory for photoionization~\cite{Fa-PhysRev-1961} 
it can be shown that these dipole phases will not affect the accuracy of the PADA method.     

With increasing photoelectron energy, the impact of electron correlation become less important and the dipole phases are quite flat.
Hence, measuring the angle-integrated or directional photoelectron spectrum may not make a significant difference, 
as we will argue in \sref{sec:secondary.results}. 
For now, we ignore the dipole phase and focus on the influence of inner-shell ionization 
and the correlation effects that are responsible for the instability of the inner-shell hole. 


\section{Inner-shell ionization}
\label{sec:core}

At 
x-ray energies, inner-shell ionization becomes possible (see \fref{fig:idea}) and it is more likely than valence or Rydberg-state ionization.
After including all possible ionization pathways, the final state after absorbing a photon in first order perturbation theory reads 
\begin{eqnarray}
  \hskip-10ex
  \label{eq:final-state}
  \ket{\Psi^{(1)}(\tau)}
  &\propto&
  \sum_{p=a,b} g_p \, e^{-i\epsilon_{p}\tau}
  \int \!dt\
  {\cal E}(t)\, e^{i\hat H_0 t} \hat z e^{-i\hat H_0 t}  \ket{\Phi^{p}_i}
  \\ \nonumber
  &\propto&
  \sum_p g_p e^{-i\epsilon_{p}\tau}
  \int\! dc\ \Big[
  z_{c,p}\,
  {\cal E}(\epsilon_c-\epsilon_{p})
  \ket{\Phi^c_i}
  +
  \sum_j
  z_{c,j} \,
  {\cal E}(\epsilon_c-\epsilon_j)
  \ket{\Phi^{cp}_{ji}}
  \Big]
  ,
\end{eqnarray}
where $\hat H_0$ is the field-free Hamiltonian with $\hat H_0 \ket{\Phi^p_i} = E^p_i \ket{\Phi^p_i}$ and $\hat H_0 \ket{\Phi^{ca}_{ji}} = E^{ac}_{ij} \ket{\Phi^{ca}_{ji}}$.
In equation~\eqref{eq:final-state}, we approximated the energy of the $2p2h$ configuration as the sum of all individual orbitals, $E^{ac}_{ij} \approx \epsilon_a + \epsilon_c - \epsilon_i -\epsilon_j$.
In this case, the photon energy is given by $\omega=\epsilon_c-\epsilon_j$ instead of $\omega= E^{ac}_{ij}- E^{a}_{i}$, which is the more general result.

The first term describes the ionization of the Rydberg electron, and the second term expresses the ionization of any other electron.
In the latter term, the ion is in an excited state, which can be written as a $1p$--$2h$ configuration, $\ket{\Phi^a_{ji}}$, where as in the first case the ion is in the ground state, $\ket{\Phi_i}$, because we required that $i$ is the an orbital in the outer-most shell.
If the orbital $j$ is an inner-shell orbital, the ionic state is not stable and hole decays. 
In \sref{sec:secondary}, we will discuss the role of secondary processes that lead to the decay of the inner-shell hole.
In this section, we assume the final ionic states are stable such that $\braket{\Phi^{c}_{i}}{\Phi^{c'}_{i'}} = \delta_{c,c'} \delta_{i,i'}$, $\braket{\Phi^{ca}_{ji}}{\Phi^{c'a'}_{j'i'}} = \delta_{c,c'} \delta_{a,a'} \delta_{j,j'} \delta_{i,i'}$, and $\braket{\Phi^{ca}_{ji}}{\Phi^{c'}_{i'}} = 0$.
%

A photoelectron spectrum shows the energy distribution of ionized electrons, $\epsilon$, 
but neither the state of the remaining ion nor the electron angular momentum---both defining the ionization channel $I$. 
Consequently, we need to add incoherently all possible ionization channels, and add coherently all pathways within the same ionization channel,
\begin{equation}
  \label{eq:pes_general}
  P(\epsilon,\tau) 
    =
  \sum_I \braket{\epsilon,I}{\Psi(\tau)}\braket{\Psi(\tau)}{\epsilon,I}
  . 
\end{equation}
Inserting Eq~\eqref{eq:final-state} in Eq.~\eqref{eq:pes_general} leads to the expression for the overall photoelectron spectrum including inner-shell ionization,
\numparts
\begin{eqnarray}
  \label{eq:pes.core}
  P(\epsilon,\tau)
  &=&
  \sum_{j\neq i} A^\textrm{\scriptsize core}_j(\epsilon) 
  + 
  \sum_{p=a,b}
  A_p(\epsilon) 
  +
  2
  B(\epsilon)  \cos\Theta(\epsilon,\tau)
  ,
  \\
  \label{eq:pes.core_coeff1}
  A^\textrm{\scriptsize core}_j(\epsilon)
  &=&
  d^2_j(\epsilon)\  |{\cal E}(\epsilon-\epsilon_j)|^2  
  ,
\end{eqnarray}
\endnumparts
where $A^\textrm{\scriptsize core}_j(\epsilon)$ is the new contribution of the inner shell, $j$, and the last two terms are identical to Eq.~\eqref{eq:pes.std}.
Note that $j$ runs over all occupied (inner and outer) orbitals except $i$ even though we refer to it as inner-shell contributions.
The signal from the inner-shell electron is not delay-dependent and contributes only to the background, 
because for each initial state, $\ket{E_{a/b}}$, the final ionic state is different, $\braket{\Phi^{a}_{ji}}{\Phi^{b}_{ji}}=\delta_{a,b}$. 
This shows that the modulation and the phase reconstruction are unaffected by inner-shell ionization when core relaxation is not taken into account. 

In contrast, all established attosecond pulse characterization methods will suffer from inner-shell ionization  
because all photoelectrons are affected by the laser field and gain delay-dependent modulations. 
Each shell will contribute delay-dependent modulations with a different energy and phase offset that be incoherently averaged.   
In the case of PROOF, for example, the overall delay-dependent modulations at a specific energy, $\epsilon$, would change to 
\begin{eqnarray}
  \label{eq:inner-PROOF}
  D_{\epsilon,i} \ {\cal E}(\epsilon-\epsilon_{i}) \ {\cal E}^*(\epsilon-\epsilon_{i}\pm\omega_L)
  &\longrightarrow&
  \sum_j D_{\epsilon,j} \
  {\cal E}(\epsilon-\epsilon_{j}) \ {\cal E}^*(\epsilon-\epsilon_{j}\pm\omega_L)
   ,
\end{eqnarray}
where $D_{\epsilon,i}$ contains all the dipole-dependent terms describing the transition from the initial to the final state.
The final modulation has still a period of $\omega_L$ but the overall phase offset cannot be directly related to a specific spectral phase difference making the pulse reconstruction more difficult. 
This makes a strong case for the PADA method, but as we allow for inner-shell relaxation in Sec.~\sref{sec:secondary}),  
we will find that there are relaxation effects that can make the secondary processes modulate also with the PADA method.

\subsection{Results}
\label{sec:core.results}

\begin{figure}[ht!]
  \centering
  \includegraphics[width=\linewidth]{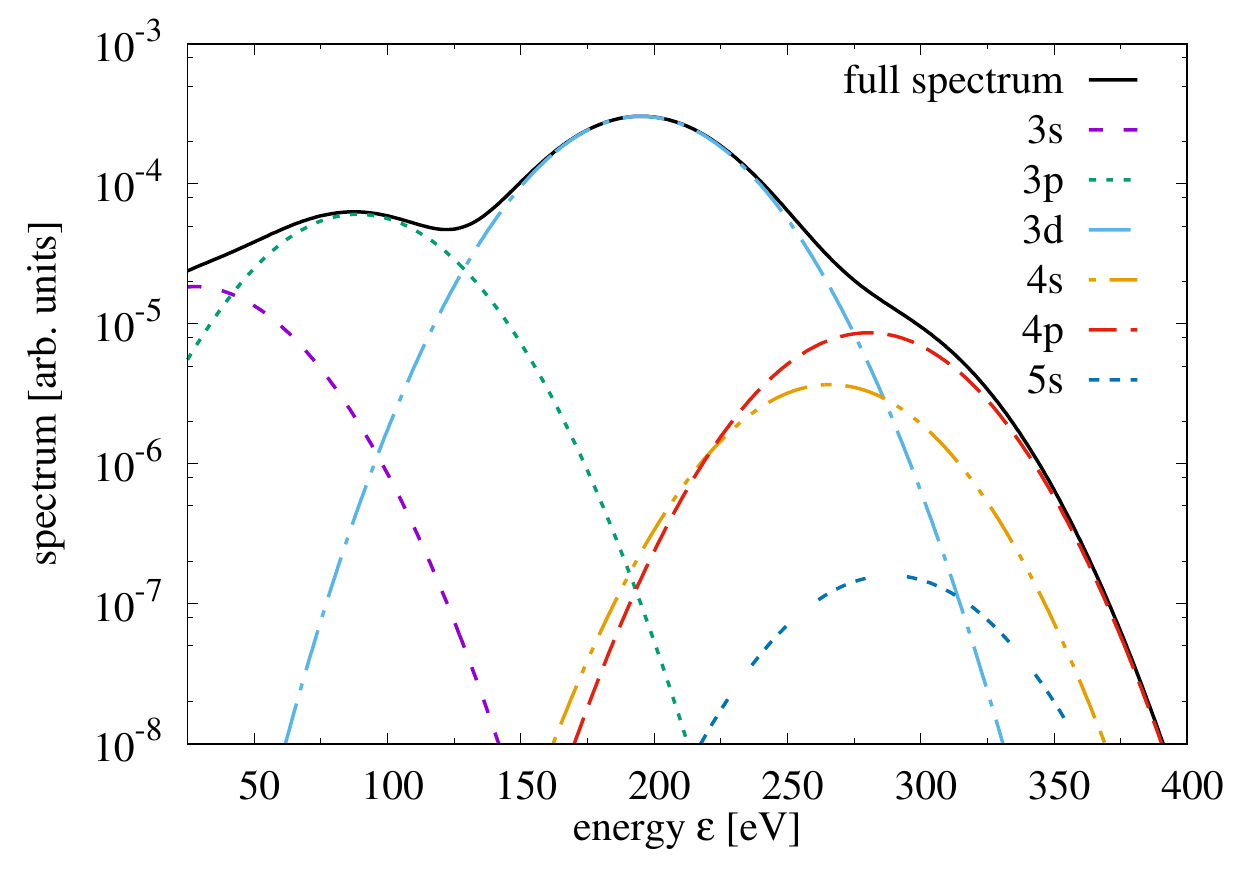}
  \caption{(color online) (a) The full and shell-resolved photoelectron spectra of krypton ionized by an 300~eV soft x-ray pulse with a spectral FWHM-width of 70~eV.
  The spectral width of the pulse is comparable to the energy separation between the electronic shells.
  }
  \label{fig:pes.gs}
\end{figure}

Figure~\ref{fig:pes.gs} shows the angle-integrated photoelectron spectrum of atomic krypton for an 300~eV attosecond pulse with FWHM spectral width of 70~eV mimicking state-of-the-art attosecond pulses~\cite{discuss-biegert}.
The contributions from the separate shells are highlighted. 
The signal of the $3d$ shell is much more dominant than that of any other shell.
Due to the broad bandwidth of the pulse, the contribution from each shell 
cannot be fully separated and we find several energy regions where two shells have the same strengths.

In general, the cross section of a Rydberg orbital (see $5s$ line in figure~\ref{fig:pes.gs}) will be at least one order of magnitude lower than that of any other occupied shell. 
The signal strength from the Rydberg wavepacket and the associated modulation strength in the spectrum are, therefore, 
a major concern for the PADA method at x-ray photon energies. 

We use Hartree-Slater to determine the orbitals, the corresponding dipole strengths, as well as the Auger and shake-up rates described in \sref{sec:auger}-\ref{sec:secondary.results}. 
Hartree-Slater has been used very successfully to describe the ionization dynamics in the x-ray regime~\cite{SoSa-PRA-2011,RuSo-NatPhot-2012}.

\begin{figure}[ht!]
  \centering
  \includegraphics[width=\linewidth]{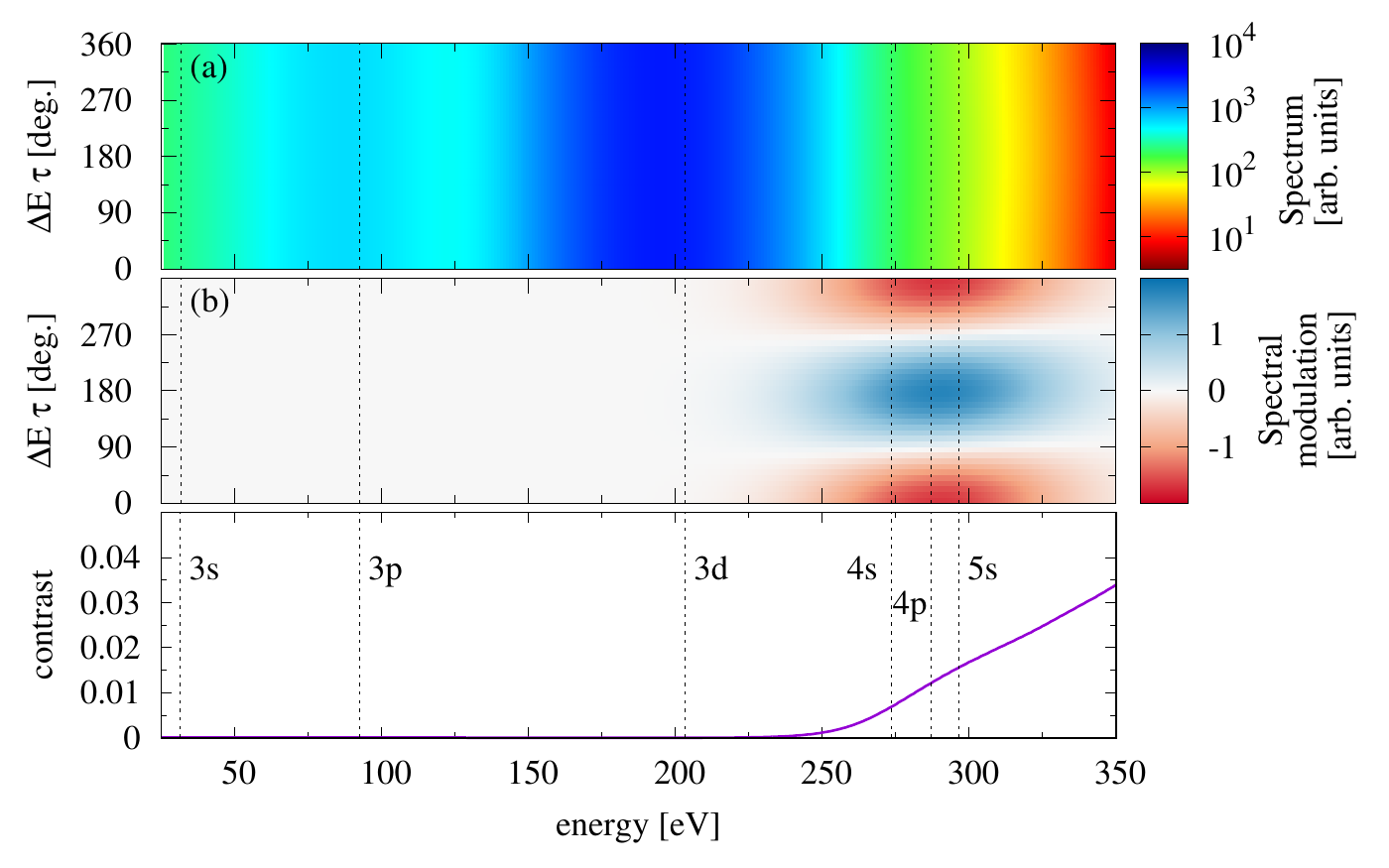}
  \caption{(color online) (a) The total and (b) the phase-dependent part of the photoelectron spectrum, $P(\epsilon,\tau)$, as the function of electron energy, $\epsilon$, and the delay, $\tau$.
  The initial state is $\ket{\Psi^{5s6s}}$---a coherent superposition of $5s$ and $6s$ in krypton.
  Only the interference between $5s$ and $6s$ is visible. 
  (c) The contrast $c(\epsilon)$ of the modulations. 
  The vertical dashed lines indicate the central position of the contribution of each shell shown in \fref{fig:pes.gs}.
  The pulse is the same as in \fref{fig:pes.gs}.
  }
  \label{fig:pes.5s6s}
\end{figure}

In \fref{fig:pes.5s6s}, the (a) total and (b) the delay-dependent modulations of the photoelectron spectrum is shown. 
The contrast $c(\epsilon)$ of the modulations are shown in (c) for a $5s$--$6s$ wavepacket in krypton, 
$\ket{\Psi^{5s6s}} = \frac{1}{\sqrt{2}}\big(\ket{\Phi^{5s}_{4p_0}} + e^{i\Delta E\, \tau} \ket{\Phi^{6s}_{4p_0}}\big)$.
The modulation contrast is defined as 
\begin{eqnarray}
  c(\epsilon)
  &=&
  \frac{\textrm{max}_\tau[P(\epsilon,\tau)]-\textrm{min}_\tau[P(\epsilon,\tau)]}{\textrm{max}_\tau[P(\epsilon,\tau)]+\textrm{min}_\tau[P(\epsilon,\tau)]}
  ,
\end{eqnarray}
where $\textrm{max}_\tau[P(\epsilon,\tau)]$ and $\textrm{min}_\tau[P(\epsilon,\tau)]$ are the maximum and minimum values of the photoelectron spectrum, $P(\epsilon,\tau)$, for a given energy, $\epsilon$, respectively.
Above 280~eV the modulations are most visible. 
In this region the photoelectron comes only from the orbitals above $4s$. 
The contrast is not one because $4s$ and $4p_{\pm1}$ orbitals contribute to a static background.
Below 280~eV the ionization is dominated by the $3d$ shell.
Below 150~eV also the $3s$ and $3p$ shells have significant contributions. 
Because there is no photoelectron delay dependence for inner-shell ionization, 
the modulations seen in \fref{fig:pes.5s6s} are exclusively due to the contribution from Rydberg electrons.  
As a result, only the energy region above 280~eV is useful for the pulse reconstruction.
The contrast goes down if the initial Rydberg state population is below 100\%. 

\begin{figure}[ht!]
  \centering
  \includegraphics[width=\linewidth]{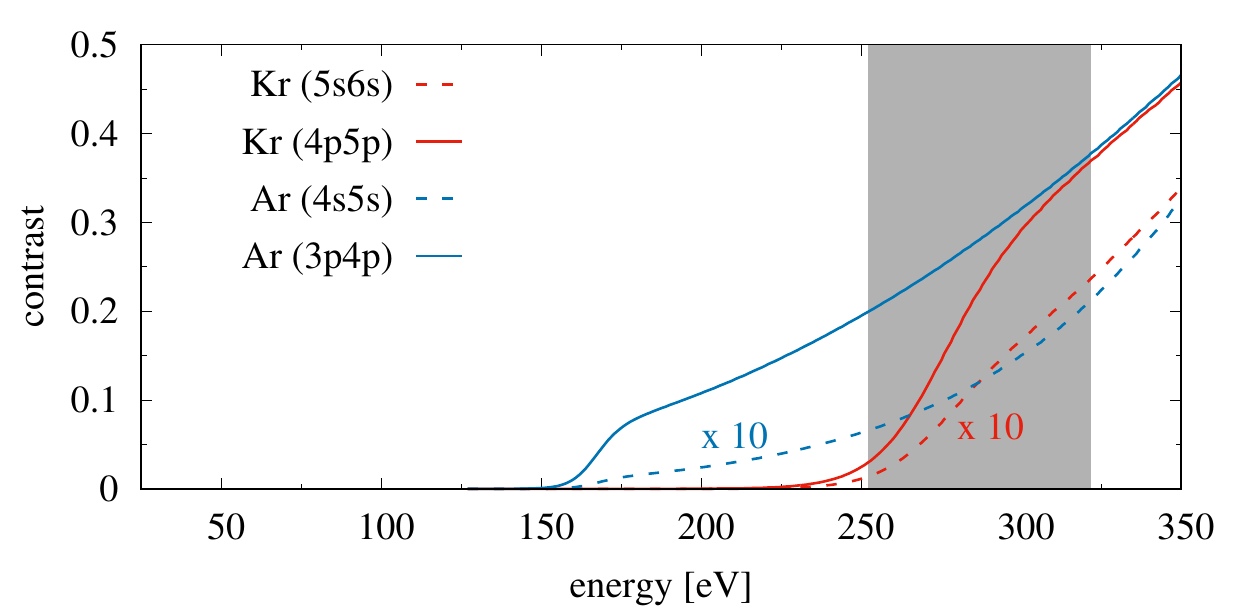}
  \caption{(color online) The contrast, $c(\epsilon)$, of the modulations in the photoelectron spectrum for (blue lines) argon and (red lines) krypton as a function of photoelectron energy, $\epsilon$.
  The contrasts for the initial wavepackets consisting of the first two $l=0$ Rydberg states (dashed lines) are multiplied by 10.
  The contrasts for the initial wavepackets consisting of the ground state and the first $l=1$ Rydberg state (solid lines) are shown as well.
  }
  \label{fig:pes.contrast}
\end{figure}

To improve the contrast, a wavepacket can be chosen as a superposition between the ground state and a Rydberg state.
For a $4p$--$5p$ wavepacket, $\ket{\Psi^{4p5p}} = \frac{1}{\sqrt{2}}\big(\ket{\Phi_0} + e^{i\Delta E\, \tau} \ket{\Phi^{5p}_{4p}(L=0)}\big)$, with total angular momentum $L=0$, the contrast is boosted by more than a factor 10 (see \fref{fig:pes.contrast}) compared to the $\ket{\Psi^{5s6s}}$ wavepacket. 
The main reason for the increase is the enhanced dipole strength of the $4p$ orbital that belongs to the ground state.  
Also in practice it is attractive to involve the ground state in the wavepacket because the modulation strength depends only linearly on the Rydberg amplitude, $g_{a/b}$ (see Eq.~\eqref{eq:wvp}), in contrast to a quadratic dependence, $g_a g_b$, for a pure Rydberg wavepacket. 
A linear scaling is favorable, because $g_{a/b}$ is normally much smaller than one for noble gas atoms as UV pulses may be required to create the initial superposition. 
A draw-back of the ground state--Rydberg wave packet is that the energy difference of the states is larger making the spectral shearing gap wider. 
This may become a problem if the test pulse has a chirp that changes by more than $2\pi$ over the energy range of the gap.

Using argon (blue lines) instead of krypton (red lines) is further beneficial since no dominant $d$-shell ionization exist in argon (see \fref{fig:pes.contrast}).
The signal from the neighboring $s$-shells is always present and fortunately relatively weak.
The next $p$-shell (i.e., $2p$-shell) is more than 230~eV away which is larger than that the spectral bandwidth of the pulse.
Going from krypton to argon show that smaller atoms are favorable for the pulse characterization. 

\section{Secondary processes}
\label{sec:secondary}

\begin{figure}[ht!]
  \centering
  \includegraphics[width=\linewidth]{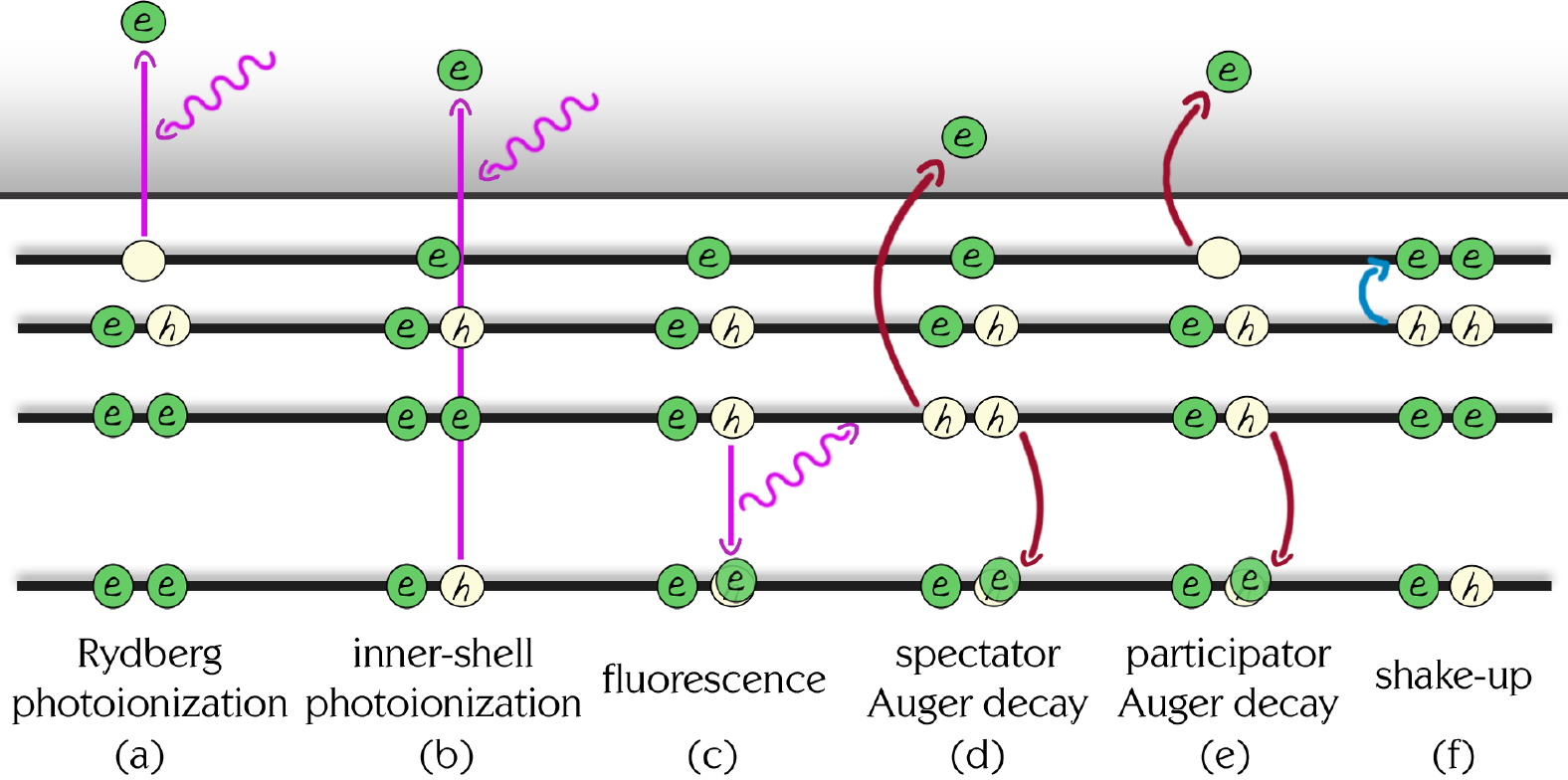}
  \caption{(color online) Sketches of (a) Rydberg and (b) inner-shell ionizations.
  (c-e) Sketches of radiative (fluorescence) and non-radiative (Auger decay) secondary processes triggered by the core hole.
  (f) Sudden creation of a core hole can lead to shake-up, where an outer electron gets promoted into a higher orbital. 
  }
  \label{fig:secondary}
\end{figure}

In \sref{sec:core}, we have seen that inner-shell ionization can limit the effective energy range that can be used to extract spectral information for the pulse reconstruction, because the static background becomes so dominant. 
Inner-shell holes are not stable and decay radiatively via fluorescence or non-radiatively via Auger decay~\cite{Si-RMP-XrayReview-1982,XDB}.
Shake-up/off is another possibility how the electronic state of the ion changes due to the sudden removal of the inner-shell electron.
These processes that follow photoionization are known as secondary processes and they are illustrated in \fref{fig:secondary}.
In this section, 
we will address the question as to how core relaxation effects can affect PADA measurements in the x-ray regime, 
thus going beyond the static inner core approximation discussed in \sref{sec:core}. 

In the case of Auger decay, the hole decays non-radiatively by moving the hole to a less bound shell and releasing the excess energy via ionization of another electron.
The final ion is doubly charged. 
In case of an excited electron in a Rydberg state, there are two types of Auger decay: spectator and participator Auger decay (see \fref{fig:secondary}d-e).
In the spectator Auger decay the Rydberg electron is not involved in the Auger process while in the participator Auger decay the Rydberg electron gets ionized.

The sudden creation of a core hole can alter the state of the electrons above due to a modified screening of the nucleus.
If an electron is promoted to another bound orbital, the process is called shake-up (see \fref{fig:secondary}f).
The case when the electron gets ionized is called shake-off (not shown in \fref{fig:secondary}).
Since the core hole does not change, it will eventually Auger decay to new channels known as satellites states\cite{Si-RMP-XrayReview-1982}.
Shake-up/off process is not very likely for electrons in initially occupied shells. 
For an electron in a Rydberg orbital, however, the probability of a shake-up event can be easily around 50\%.
Similarly to the Auger decay, there exist spectator and participator processes for shake-up/off and also fluorescence depending whether or not the Rydberg electron is involved in the process.

Secondary processes lead to singly-charged or doubly-charged ions.
It is even possible that the same final ionic state can be reached from different initial Rydberg excitations. 
Consequently, inner-shell ionization can lead to delay-dependent interference terms affecting the pulse reconstruction.
All established pulse characterization methods are affected by secondary processes. 
Next, we study the detailed influence of fluorescence, Auger decay, and shake-up/off on the PADA method.


\subsection{Fluorescence}
\label{sec:fluorescence}

In the case of fluorescence, an electron from a higher-lying shell fills the inner-shell hole, $j$, and emits a photon (see \fref{fig:secondary}c),
\
\begin{eqnarray}
  \label{eq:fluo}
  \ket{\Phi^a_{ji};0}
  \stackrel{\textrm{fluorescence}}{\longrightarrow}
  \sum_{j'}
  d_{j,j'}\ket{\Phi^a_{j'i};\omega_{j'j}}
  +
  d_{j,a}
  \ket{\Phi_{i};\omega_{aj}}
  .
  \end{eqnarray}
The composite state $\ket{\Phi_{i};\omega_{aj}}$ includes the ion, $\ket{\Phi_{i}}$, and the fluorescence photon, $\ket{\omega_{aj}}$, with energy $\omega_{aj}=\epsilon_{a}-\epsilon_j$.
Initially no fluorescence photon is present, which is expressed by $\ket{0}$.
The dipole transition strength is given by $d_{p,q}=\bra{p}\hat d\ket{q}$.
The final state is characterized by the ion and the emitted photon.
In the first term in equation~\eqref{eq:fluo} the Rydberg electron is a spectator leading to a $1p$--$2h$ configuration in the ion, whereas in the second term the Rydberg electron participates in the fluorescence process leading to a $1h$ configuration.
After including the fluorescence decay, the overlap between the core-hole states reads,
\begin{eqnarray}
  \label{eq:fluores.olap}
  \hskip-10ex
  \braket{\Phi^{a}_{ji};0}{\Phi^{b}_{ji};0}
  &\longrightarrow&
  \sum_{j',j''}
  d_{j'j} d_{j''j}
  \braket{\Phi^{a}_{j'i}}{\Phi^{b}_{j''i}}
  \braket{\omega_{j'j}}{\omega_{j''j}}
  +
  d_{j,a}d_{j,b} \braket{\Phi_{i}}{\Phi_{i}}
  \braket{\omega_{a j}}{\omega_{b j}}
  ,
\end{eqnarray}
which has a contribution from the spectator (first term) and the participator (second term) fluorescence decay.
If $\ket{\Phi^{a}_{j'i}}$ is a stable configuration (the inverse lifetime is much smaller than the energy separation between the two Rydberg states), 
then we can write $\braket{\Phi^{a}_{j'i}}{\Phi^{b}_{j''i}}=\delta_{a,b}\delta_{j',j''}$.
The $\delta_{a,b}$ term ensures that the same final ionic state cannot be reached from different initial states. 
%
For the participator event [second term in equation~\eqref{eq:fluores.olap}], the final ionic state is the same and stable because $i$ represents an outer shell.
Here, the photonic overlap, $\braket{\omega_{aj}}{\omega_{bj}}=\delta_{a,b}$, enforces that both Rydberg states have to be the same, which means also the participator channel does not affect the pulse characterization. 
%
In more detail, the orthogonality conditions used above are only true if the {\it total} decay rate of the hole $j$ 
is much smaller than the energy separation between the Rydberg states. 
The total decay rate for core holes should include both fluorescence and Auger decay. 
While fluorescence rates in the soft x-ray regime may be negligible, the corresponding Auger decay rates are greater, typically around 100~meV. 
This corresponds to typical energy separations of high Rydberg states and, therefore, 
the fluorescence decay can affect the pulse characterization procedure due to the fast depletion of the core hole.
At hard x-rays the situation is reversed as fluorescence decay dominates over Auger decay.

\subsection{Auger decay}
\label{sec:auger}

In the case of an Auger event, a core hole $j$ decays via electron-electron interaction by filling the hole with an electron from a higher-lying orbital, $j_{1}$, and giving the excess energy to another electron in orbital $j_{2}$, which has now enough energy to escape the atom.
In terms of CI coefficients, it reads
\numparts
\begin{eqnarray}
  \label{eq:state.auger}  
  \ket{\Phi^a_{ji}}
  &\stackrel{\textrm{Auger} }{\longrightarrow}&
  \sum_{j_1,j_2} \int dc\  
  \beta^{c}_{j;j_1,j_2} 
  \ket{\Phi^{ca}_{j_1j_2i}}
  +
  \sum_{j_1}  \int dc\  
  \beta^{c}_{j;j_1,a} 
  \ket{\Phi^{c}_{j_1i}}
  ,
  \\
  \label{eq:state.auger.coeff}
  \beta^c_{j;j_1,p} 
  &=&
  \frac{w}{\sqrt{ (\epsilon_c-E^j_{j_1 p})^2 + \Gamma_j^2/4}}
  ,
\end{eqnarray}
\endnumparts
where $c$ represents the Auger electron, $\epsilon_c$ is the energy of the Auger electron, and $\Gamma_j$ is the total decay rate of the $j$ hole.
The energy difference, $E^j_{j_1 p}$, between the initial state $\ket{\Phi^a_{ji}}$ and the final ionic state $\ket{\Phi^a_{j_1 j_2 i}}$ ($p=j_2$) and $\ket{\Phi_{j_1 i}}$ ($p=a$), respectively, has generally a weak $i$ and $a$ dependence.
In an independent particle picture, the energy difference  $E^j_{j_1 p}=\epsilon_{j_1}+\epsilon_{p}-\epsilon_{j}$ does truly depend only on $j, j_1,$ and $p$.
The Auger strength, $w$, is given for the spectator (participator) decay by $w=V_{j;j_1,j_2;c} =\bra{\Phi^{ca}_{j_1j_2 i}}r^{-1}_{12}\ket{\Phi^a_{ji}}$ ($w=V_{j;j_1,a;c} =\bra{\Phi^{c}_{j_1i}}r^{-1}_{12}\ket{\Phi^a_{ji}}$).
A detailed derivation of the $\beta$ coefficients is given in the \ref{app:auger-lineshape}.
The new overlap reads,
\begin{eqnarray}
  \label{eq:auger.olap}
  \braket{\Phi^{a}_{ji}}{\Phi^{b}_{ji}}
  &\longrightarrow&
  \sum_{j_1,j_2}  \int dc\  
  |\beta^{c}_{j,j_1,j_2}|^2
  \delta_{a,b}
  +
  \sum_{j_1}  \int dc\  
  [\beta^{c}_{j,j_1,a}]^*
  \beta^{c}_{j,j_1,b}
  ,
\end{eqnarray}
where the first (second) term arises from the spectator (participator) Auger decay.
Similarly to the fluorescence case, the spectator decay leads for different initial states to different final states contributing to a delay-independent background. 
For the participator Auger decay (second term), the final ionic state is the same, $\ket{\Phi_{j_1i}}$, and the orthogonality between the two states holds as long as the Auger electrons are energetically distinguishable, i.e., $|E^j_{j_1 a}-E^j_{j_1 b}|\gg\Gamma_j$  such that $\int dc\ [\beta^{c}_{j,j_1,a}]^* \beta^{c}_{j,j_1,b} \sim \delta_{a,b}$.
The energy separation between the lowest Rydberg states is around 1--2~eV and only deep core holes decay fast enough ($\lesssim 300$~as) to bridge this energy gap. 
In krypton, for instance, core holes below the $3d$ shell are required.

The photoelectron spectrum\footnote{We ignore the contribution of the Auger electron to the electron spectrum because it contributes only at very specific energies to the spectrum.} after including the effect of Auger decay is found by substituting equation~\eqref{eq:final-state} into equation~\eqref{eq:pes_general} and replacing the overlaps between the ionic states with the expressions in equation~\eqref{eq:auger.olap}.
The inner-shell contributions to the photoelectron spectrum, which appear additionally to the ones in equation~\eqref{eq:pes.std}, are
\begin{eqnarray}
  \label{eq:pes.auger.simple}
  \hskip-12ex
  P^\textrm{\scriptsize Auger}(\epsilon,\tau)
  &=&
  \sum_{j,j_1}
  \Big[
  \sum_{j_2} A^\textrm{\scriptsize Auger}_{j;j_1,j_2}(\epsilon) 
  + 
  \sum_{p=a,b} A^\textrm{\scriptsize Auger}_{j;j_1,p}(\epsilon) 
  +
  2
  B^\textrm{\scriptsize Auger}_{j;j_1}(\epsilon)  \cos\Theta^\textrm{\scriptsize Auger}_{j;j_1}(\epsilon,\tau)
  \Big]
  ,
  \nonumber\\
\end{eqnarray}
where $A^\textrm{\scriptsize Auger}$ are the static contributions, $B^\textrm{\scriptsize Auger}$ the strength of modulation from the participator Auger decay, and $\Theta^\textrm{\scriptsize Auger}$ contains the phase dependence of the modulation.
The coefficients read
\numparts
\begin{eqnarray}
  \label{eq:pes.auger.simple.coeff1}
  \hskip-10ex
  A^\textrm{\scriptsize Auger}_{j;j_1,j}(\epsilon) 
  &=&
  d^2_j(\epsilon)  \frac{2\pi|V_{j;j_1,j}|^2}{\Gamma_j} \
  \bar{\cal E }^2_{j;0}(\epsilon-\epsilon_j)
  ,
  \\
  \label{eq:pes.auger.simple.coeff1.1}
  \hskip-10ex
  A^\textrm{\scriptsize Auger}_{j;j_1,p}(\epsilon) 
  &=&
  g_p^2\, d^2_j(\epsilon)  \frac{2\pi|V_{j;j_1,p}|^2}{\Gamma_j} \
  \bar{\cal E }^2_{j;0}(\epsilon-\epsilon_j)
  ,
  \\
  \label{eq:pes.auger.simple.coeff2}
  \hskip-10ex
  B^\textrm{\scriptsize Auger}_{j;j_1}(\epsilon) 
  &=&
  g_a g_b\,
  d^2_j(\epsilon)
  \frac{2\pi|V_{j;j_1,a} V_{j;j_1,b}|}{\Gamma_j} \
  \bar{\cal E}^2_{j;\Delta E}(\epsilon-\epsilon_j) 
  ,
  \\
  \label{eq:pes.auger.simple.coeff3}
  \hskip-10ex
  \Theta^\textrm{\scriptsize Auger}_{j;j_1}(\epsilon,\tau) 
  &=&
  \textrm{arg}\Big( \Delta E \tau + [\phi(\epsilon-\epsilon_j-\Delta E/2)-\phi(\epsilon-\epsilon_j+\Delta E/2)] \Big)
  ,
\end{eqnarray}
\endnumparts
where $\Delta E=\epsilon_a-\epsilon_b$ is the energy difference between the Rydberg states in the wavepacket [see equation~\eqref{eq:wvp}], the coupling strength $V_{j;j_1,p}=\sum_c V_{j;j_1,p;c}$ is evaluated at resonant energy $\epsilon_c=E^j_{j_1,p}$, and 
\begin{eqnarray}
  \label{eq:pes.auger.pulse-avrg}
  \bar{\cal E}^2_{j;\Delta E}(\omega) 
  &=&
  \frac{\Gamma_j}{2\pi}\int d\epsilon
  \frac{|{\cal E}(\omega+\epsilon-\Delta E/2) {\cal E}(\omega+\epsilon+\Delta E/2)|}
  {
  \sqrt{
    \big[(\epsilon-\Delta E/2)^2 + \Gamma_j^2/4 \big]
    \big[(\epsilon+\Delta E/2)^2 + \Gamma_j^2/4  \big]
    }
  }
  ,
\end{eqnarray}
with the limit $\bar{\cal E}^2_{j;\Delta E}(\omega)\stackrel{\Gamma_j\rightarrow 0}{\longrightarrow} |{\cal E}^2(\omega)|^2$ iff $\Delta E=0$.
Equation~\eqref{eq:pes.auger.pulse-avrg} shows the energy uncertainty of the Auger electron, which is given by $\Gamma_j$, results in an uncertainty in absorbed photon energy, which affects which spectral phases are probed at a given photoelectron energy, $\epsilon$.
To arrive at Eqs.~\eqref{eq:pes.auger.simple}--\eqref{eq:pes.auger.simple.coeff3} we made two approximations: (1) the coupling strength $V_{j;j_1,p:c}$~\cite{Fa-PhysRev-1961} and (2) the difference in the spectral phase, $\phi(\epsilon-\epsilon_j-\Delta E/2)-\phi(\epsilon-\epsilon_j+\Delta E/2)$, do not vary across the resonance.
The exact result without approximations is given in \sref{app:auger}.

%
The relative strength of the modulations compared to the static background is approximately given by $|V_{j;j_1,a} V_{j;j_1,b}|/\Gamma_j$, where $2\pi| V_{j;j_1,a/b}|^2$ is the participator Auger decay rate for the $a/b$ Rydberg state.
Participator Auger decays for the lowest Rydberg states (with a core hole) are orders of magnitude less likely than spectator Auger decays~\cite{note.HS-based}.
Consequently, the modulation strength due to the Auger decay is quite weak. 

For example, the lifetime of the $3d$ hole in krypton is $\sim14$~fs corresponding to $\Gamma_{3d}=46$~meV.
The energy separation between the lowest Rydberg states is around 1--2~eV and much larger than $\Gamma_{3d}$ such that no delay-dependent modulations are expected for a $3d$ inner-shell ionization.
For an $3s$ hole in krypton, the scenario changes because $\Gamma_{3s}\approx 7$~eV~\cite{note.HS-based} and exceeds the energy separation between Rydberg states.
We see whether or not interferences due to Auger decay occur depends on the hole and the Rydberg states involved in the wavepacket.

\subsection{Shake-up}
\label{sec:shakeup}

Shake-up occurs when a core electron is suddenly removed, and the remaining electrons in the system rearrange accordingly.  
The main effect of the electron removal is the reduced Coulomb screening of the nucleus, which results in a contraction of the orbitals.
Consequently, an electron in an initial $(n,l,m)$-orbital may end up in a $(n',l,m)$-orbital (shake-up) or even in a $(\epsilon,l,m)$-continuum state (shake-off).
While the angular characteristics of the electron do not change due to a modified central potential, the radial orbitals will be contracted.  
Since shake-up is much more likely than shake-off, we focus our discussion on shake-up events. 
The description for shake-off is very similar to shake-up just that sums over bound Rydberg states have to be replaced with integrals over continuum states.
A core-hole configuration after shake-up reads
\begin{eqnarray}
  \label{eq:state.shake-up}
  \ket{\Phi^a_{j} }
  &&\stackrel{\textrm{shake-up} }{\longrightarrow}\
  \sum_{p'} \gamma^j_{p',j_1} 
  \ket{\Phi^{a'p'}_{j'_1 j'}}
  +
  \sum_{j_1,p'}
  \gamma^j_{p',a} 
  \ket{\Phi^{p'}_{j'}}
  ,
\end{eqnarray}
where we use primes to indicate the contracted orbitals.
The shake-up amplitudes are the overlap of the initial orbital with newly contracted orbitals, $\gamma^j_{p',q}=\braket{p'}{q}$, where the dependence of the hole $j$ is implicit as it defines how the orbital $p'$ is contracted.
For delocalized orbitals such as Rydberg states the hole dependence is weak because highly excited states just see that an electron is missing but the exact shape of the (localized) core hole does not matter.
The sum over $j_1$ runs over orbitals less bound than the $j$ hole because less bound orbitals are much stronger affected by the modified core screening than deeper bound ones.
Also energy conservation ensures that only shells above the core hole are affected.

It is possible to have two or more electrons be shaken-up simultaneously.
The shake-up probability is much lower for multiple shake-ups rendering them less likely.
Therefore, we focus on the leading order where only one electron is shaken-up.

Equation~\eqref{eq:state.shake-up} contains a spectator (first term) and a participator (second term) shake-up event.
Only the participator process can lead to a delay-dependent interference because the spectator process leads to distinctly different ionic states (similarly to fluorescence and Auger decay).
In contrast to the Auger decay, participator shake-up is more likely than the spectator shake-up because a Rydberg electron is much more likely to change to a neighboring Rydberg state than an electron from an occupied orbital gets shaken-up into a Rydberg state.
Consequently, we focus on the dominant participator event.


The shake-up contributions to the photoelectron spectrum, which appear additionally to the ones in equation~\eqref{eq:pes.std}, are
\begin{eqnarray}
  \label{eq:pes.shakeup}
  P^\textrm{\scriptsize shake}(\epsilon,\tau)
  &=&
  \sum_{j\neq i} \Big[
  \sum_{p=a,b}
  A^\textrm{\scriptsize shake}_{j;p}(\epsilon) 
  +
  \sum_{p'}
  B^\textrm{\scriptsize shake}_{j;p'}(\epsilon)\  \cos\Theta^\textrm{\scriptsize shake}_{p',j}(\epsilon)
  \Big]
  ,
\end{eqnarray}
with
\numparts
\begin{eqnarray}
  \hskip-5ex
  \label{eq:pes.shake_coeff1}
  A^\textrm{\scriptsize shake}_{j,a}(\epsilon)
  &=&
  g_a^2\,  d^2_j(\epsilon)
    \sum_{p'} [\gamma^j_{p',a}]^2\ |{\cal E}(\epsilon+E'^{p}_{i,j}-E^{a}_{i})|^2  
  ,
  \\
  \hskip-5ex
  \label{eq:pes.shake_coeff2}
  B^\textrm{\scriptsize shake}_{j,p'}(\epsilon)
  &=&  
  g_a g_b\,
  d^2_j(\epsilon)\   
  \gamma^j_{p',a}\, \gamma^j_{p',b} \ 
  |{\cal E}(\epsilon+E'^{p}_{i,j}-E^{a}_{i})\, {\cal E}(\epsilon+E'^{p}_{i,j}-E^{b}_{i})|
  ,
  \\
  \hskip-5ex
  \label{eq:pes.shake_coeff3}
  \Theta^\textrm{\scriptsize shake}_{j,p'}(\epsilon)
  &=&
  \textrm{arg}\Big(
  \Delta E\,\tau + [\phi(\epsilon+E'^{p}_{i,j}-E^{b}_{i})-\phi(\epsilon+E'^{p}_{i,j}-E^{a}_{i})]
  \Big)
  ,
\end{eqnarray}
\endnumparts
where $E'^{p}_{i,j}$ is final energy of the ion after contraction. 
The ion energy is calculated by performing a self-consistent mean-field calculation based on Hartree-Slater with enforcing a hole in the $j$ orbital~\cite{SoSa-PRA-2011}.
Shake-up does not affect the phase, $\Theta^\textrm{\scriptsize shake}$, because the shake-up transition is an overlap between bound states, which is always real (as long as other correlation effects are ignored).


In krypton, the probability of shake-up from $5p$ to $6p$ due to an $3s$ hole is $[\gamma^{3s}_{6p',5p}]^2 \approx 0.30$.
The probability of staying in the $5p$ orbital is 69\%, and the probability to reach any other orbitals is around 1\%.
The probability of shake-up from $4p$ to $5p$ due to a $3s$ hole is 1\% with a 98\% probability to stay in $4p$.
We see shake-up is much less likely for electrons in initially occupied orbitals than for Rydberg electrons.
Furthermore, we find shake-up is happening predominantly to the next higher lying orbital.


\subsection{Results}
\label{sec:secondary.results}

After we have formulated the implication of fluorescence, Auger decay, and shake-up to the photoelectron spectrum, we turn to an explicit example.
In \sref{sec:core.results}, we found that using a wavepacket that involves the ground state is most desirable for seeing the delay-dependent modulations.
In the case of krypton, we pick $\ket{\Psi^{4p5p}} = \ket{\Phi_0} + e^{i \Delta E\, \tau} \ket{\Phi^{5p}_{4p}(L=0)}$, where both states have overall angular momentum $L=0$.
The energy difference between the two states is 11.6~eV (10.6~eV within Hartree-Slater).
Auger decay can only influence the delay-dependent modulations if a hole decays in less than 50~as.
For krypton, the fastest hole decay is the $3s$ hole with a lifetime of 96~as~\cite{note.HS-based}.
A hole in the dominant $3d$ shell decays within 14~fs, which would not even impact a $5p$--$6p$ Rydberg wavepacket.
Consequently, we do not need to worry about Auger decay or fluorescence as they affect only slightly the delay-independent background.

The only secondary effect we need to consider is shake-up, with the dominant process being a $4p$ electron promoted to $5p$.
As mentioned in \sref{sec:shakeup}, this probability is 1\% and greatly reduced from the shake-up probability of a Rydberg electron ($\gtrsim 30\%$).
Here, we already see that by choosing a suitable initial wavepacket, we can control the influence of secondary processes on the pulse reconstruction.

\begin{figure}[ht!]
  \centering
  \includegraphics[width=\linewidth]{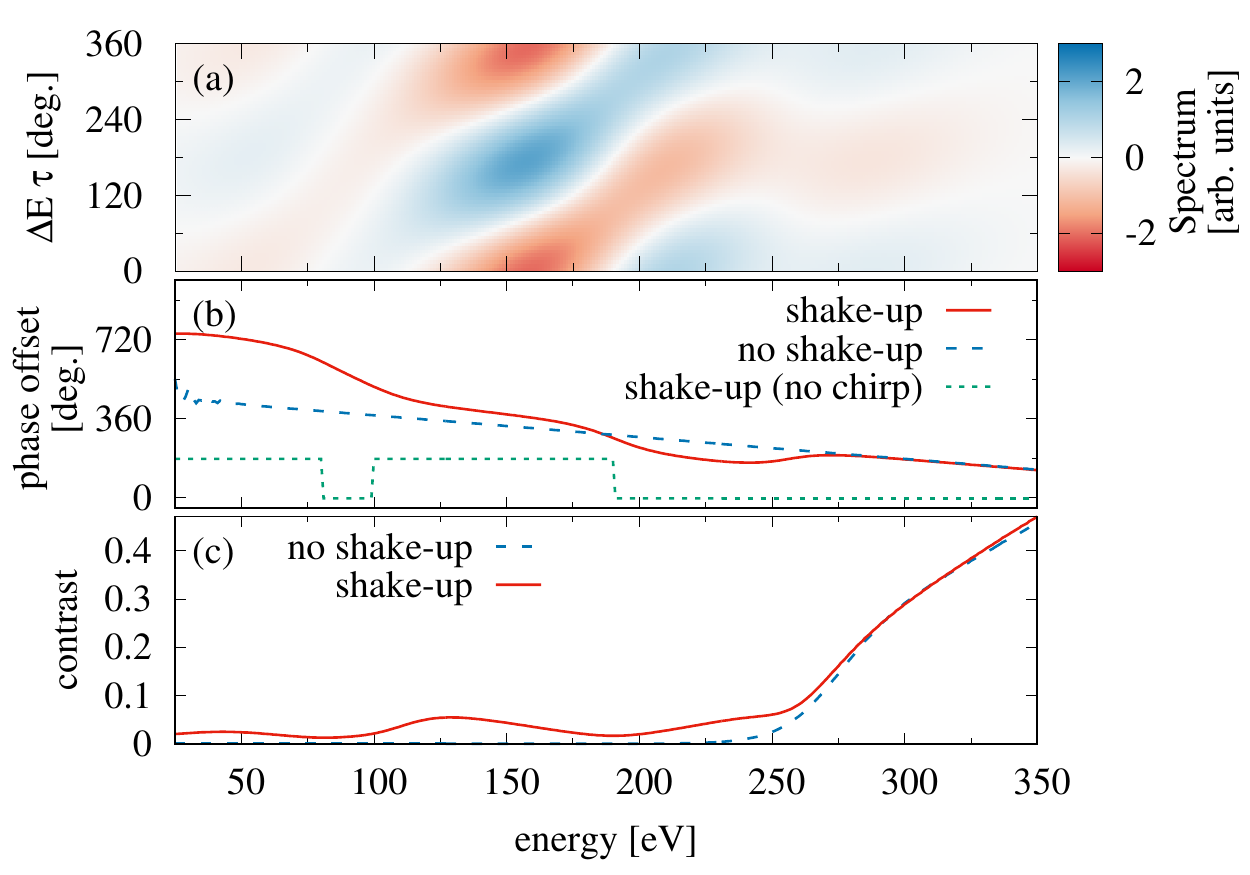}
  \caption{(color online) (a) The photoelectron spectrum, $P(\epsilon,\tau)$, for krypton including shake-up processes when ionizing core shells. 
  The initial wavepacket is $\ket{\Psi^{4p5p}}$.
  (b) The phase offset and (c) the contrast of the modulations with (solid red line) and without (dashed blue line) shake-up processes.
  In (b) also the phase offset for a Fourier-limited pulse (green dotted line) is shown.
  }
  \label{fig:shakeup}
\end{figure}

Even though the shake-up probability of $4p$ to $5p$ is only 1\%, the main modulation signal due to superposition of the outer electron is weak as well because the reduced dipole moment of $5p$ is more than 100 times smaller than the one of $3d$.
In \fref{fig:shakeup}(a), the delay-dependent part of the photoelectron spectrum of krypton is shown with the initial wavepacket $\ket{\Psi^{4p5p}}$.
The pulse is the same as in \sref{sec:core}, which was a linearly chirped Gaussian pulse with central frequency 300~eV and 70~eV bandwidth.
It is clearly visible that the slope of the modulations [see \fref{fig:shakeup}(b)] is not linear, as we would expect from a linearly chirped pulse (blue dashed line), and shows a complex dependence that does not directly reflect the spectral phases of the pulse.
As comparison, the phase offset for a Fourier-limited pulse with no chirp (green dotted line) is shown as well, and experiences $180^\circ$ jumps due shake-up (see below for explanation).


The PADA modulation coincides with the expected result for $\epsilon>280$~eV, 
because the contributions at these photoelectron energies are dominanted by the outer shell ionization. 
The contrast is roughly $1/3$ due to background photoionization of the remaining $4p$ electrons, while the contribution from $4s$ electrons is smaller. 
Around 280~eV the signal from the $4p$ and $3d$ shells are comparable and lead to deviations as the $3d$ and $4p$ shell probes different spectral components.
The 280~eV energy position is specific to the 70~eV broad pulse with $\omega=300$~eV. 
For different pulse parameters, this position moves accordingly.
At 225~eV the $3d$ shell dominates and we observe that the PADA modulation is shifted down to roughly the same value as in the outer-shell region at $\epsilon>280$~eV.  

The phase change around 100~eV and 200~eV is due to a sign flip in the $B$ coefficient [see equation~\eqref{eq:pes.shake_coeff2}] and more precisely in a sign change in the shake-up amplitude $\gamma_{5p',4p}$ compared to $\gamma_{4p',5p}$ due to a $3p$ and $3d$ hole~\cite{note-shakeup}, respectively.
The sign flip in the shake-up amplitudes is most visible for the Fourier-limited pulse (green dotted line) with abrupt $180^\circ$ jumps as one shake-up channel becomes more dominant than another.
The phase change due to the transition from the $3d$ shell to the $3p$ shell is much weaker than the shake-up effect at 100~eV.
For Fourier-limited pulses, contribution from different shells do not lead to phase changes because the spectral phase of the pulse is energy independent, ($\phi(\epsilon)=\textrm{const.})$.

Overall, \fref{fig:shakeup}(b) shows the phase dependence in the spectrum are not trivially connected to the spectral phases of the pulse.
The delay-dependent modulations due to shake-up are visible at much lower electron energies.
This may potentially help the reconstruction because a larger energy region can be used to analyze the phases, even though the phase offset behavior is connected to the spectral phases in a more complicated manner.

To mitigate shake-up, a wavepackets not consisting of a coherent superposition between neighboring orbitals could be chosen.
In this case, the shake-up probability would be greatly reduced. 
Another approach to mitigate the influence of shake-up is by choosing a wavepacket with orbitals of different parity or angular momentum.
This is advantageous because shake-up is not likely to change the angular character of the electron. 
When choosing a wavepacket containing different angular momenta, it is necessary to measure the directional photoelectron spectrum to be able to see delay-dependent interferences~\cite{PaDa-PRA-2016}.

\begin{figure}[ht!]
  \centering
  \includegraphics[width=\linewidth]{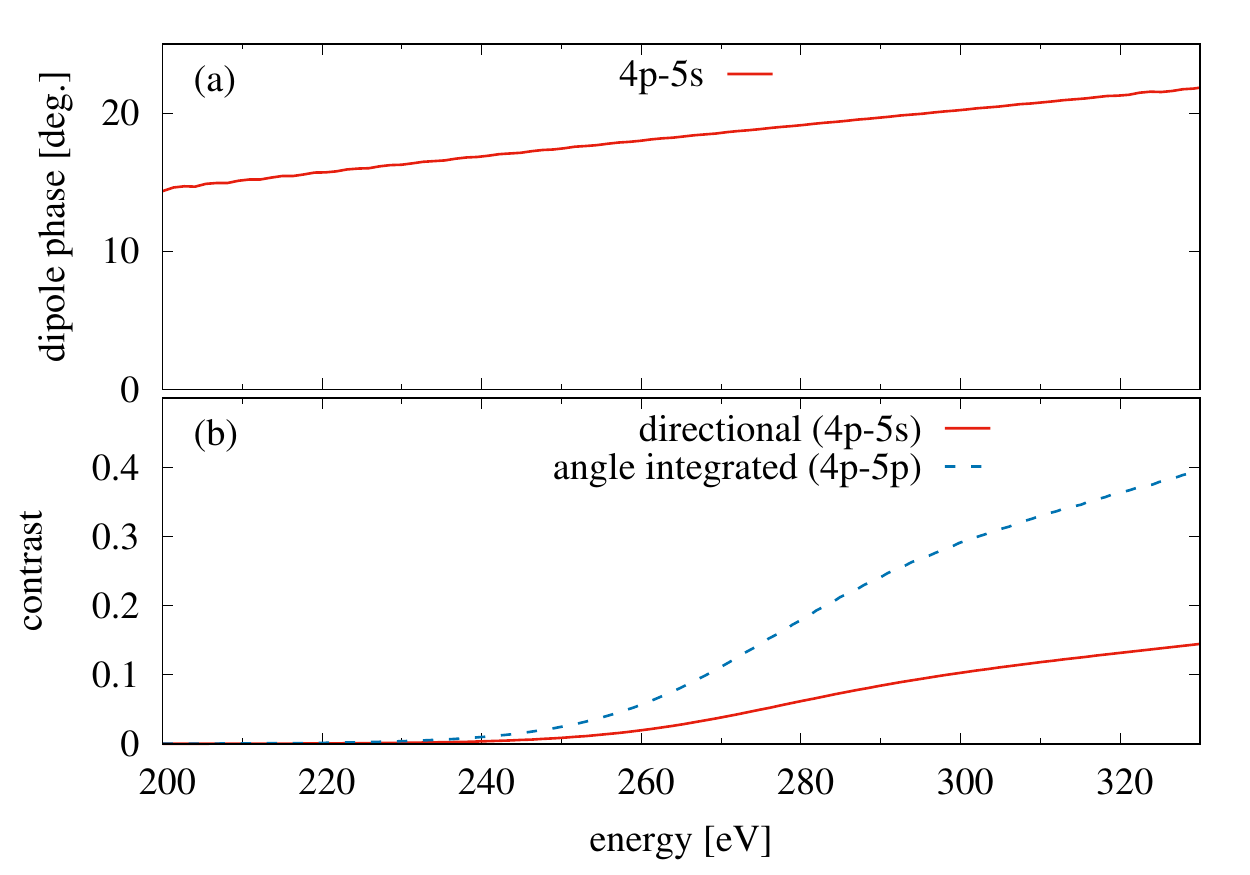}
  \caption{(color online) (a) The phase offset of the modulation of a directional photoelectron spectrum in krypton. 
  The initial state is $\ket{\Psi^{4p5s}}\propto \ket{\Phi_0} + e^{i\Delta E\, \tau} \ket{\Phi_{4p_0}^{5s}}$.
  (b) The contrast of the modulations for the directional photoelectron spectrum with the initial state $\ket{\Psi^{4p5s}}$ (solid red line) and for the angled integrated photoelectron with the initial state $\ket{\Psi^{4p5p}}$ (dashed blue line).  
  }
  \label{fig:direct}
\end{figure}

The drawback of the directional photoelectron spectrum is that the dipole phase reenters in Eq.~\eqref{eq:pes.std_coeff3}.
In the soft x-ray regime, however, the dipole phase is quite flat in contrast to the UV regime. 
In \fref{fig:direct}(a), the phase offset of the modulation for a directional photoelectron spectrum is shown for a Fourier-limited pulse with the initial wavepacket $\ket{\Psi^{4p5s}}\propto \ket{\Phi_0} +  e^{i\Delta E\, \tau} \ket{\Phi_{4p_0}^{5s}}$.
Only the linear slope of 0.079~deg./eV corresponding to a chirp of $c=0.12$~as/eV is of interest.
By using the relation~\cite{PaDa-PRA-2016}, $\tau_\textrm{\scriptsize crit}[\textrm{as}]=42.7\sqrt{c[\textrm{as/eV}]}$, we find the dipole phase is only problematic for pulses shorter than 13~as.

In \fref{fig:direct}(b), the contrast of the modulations for the $\ket{\Psi^{4p5s}}$ wavepacket (solid red line) in the directional photoelectron spectrum is compared with the contrast of the same-parity $\ket{\Psi^{4p5p}}$ wavepacket (blue dashed line) in the angle-integrated photoelectron spectrum.
The contrast for $\ket{\Psi^{4p5s}}$ is roughly a factor 3 weaker than for $\ket{\Psi^{4p5p}}$ because all three $m=\pm1,0$ components contribute in $\ket{\Phi^{4p5p}}$ but only $m=0$ in $\ket{\Psi^{4p5s}}$.
Furthermore, it is experimentally attractive to create $\ket{\Psi^{4p5s}}$ because it requires only a one-photon transition and not a two-photon transition. 

\section{Conclusion}
\label{sec:conclusion}

Attosecond x-ray pulses will bring core-hole spectroscopy into the attosecond regime, and will open up new ways to trigger and to probe ultrafast electronic and nuclear motions.
Knowing the exact shape and the spectral phases is particularly important when using the attosecond pulse to launch electronic wavepackets~\cite{KlVr-Science-2006} and studying sub-cycle dynamics~\cite{WiGo-Science-2011}.
However, new challenges arise for the characterization of these pulses as inner-shell ionization becomes dominant and secondary processes introduce new features in the photoelectron spectrum.

In traditional attosecond pulse characterization techniques each ionization channel creates new modulations in the spectrum that need to be 
disentangled to retrieve the spectral phase of the test pulse. 
With Pulse Analysis by Delayed Absorption (PADA), 
a recently proposed method that is based on ionization of bound wavepackets \cite{PaDa-PRA-2016},
we have shown that inner-shell ionization contributes only to a static background signal.   
The contrast issue was studied  because cross sections of inner orbitals are typically larger than of valence and Rydberg orbitals at x-ray energies. 

Secondary processes such as fluorescence, Auger decay, and shake-up lead to further averaging over spectral phases.
The impact of low contrast and secondary processes can be mitigated by choosing an appropriate wavefunction.
We find lighter atoms are generally more favorable as secondary processes are reduced and the energy separation between neighboring shells is larger. 

Thanks to the flexibility of choosing an appropriate wavepacket, future improvements to our approach can be made.
We expect that the usage of a hole wavepacket instead of a Rydberg wavepacket can overcome some of the visibility challenges that exist for Rydberg wavepackets.
Characterizing the pulse by analyzing the transient absorption signal and not the photoelectron spectrum is another interesting possibility that deserved further attention.

Overall, the PADA approach offers high flexibility and can eliminate the unwanted side effects due to inner-shell ionization making it a promising and reliable pulse characterization method for attosecond x-ray pulses.

\ack
S.P. is funded by the Alexander von Humboldt Foundation and by the NSF through a grant to ITAMP.
J.M.D. is funded by the Swedish Research Council, Grant No. 2014-3724.

\appendix

\section{Derivation of the Auger $\beta$ coefficient}
\label{app:auger-lineshape}

The energy distribution of the Auger electron, which is given by $|\beta(\epsilon)|^2$, leads to a Lorentzian distribution.
To arrive at equation~\eqref{eq:state.auger.coeff}, it is helpful to project the initial core-hole state and the final doubly ionized state on the exact eigenstates of this multi-channel problem, which are given by~\cite{Fa-PhysRev-1961}
\begin{equation}
  \label{eq:fano.state}
  \ket{E}
  =
  d_0 \ket{\Phi^a_{ji}}
  +
  \sum_{j_1,j_2} \int d\epsilon_c \ d_{j_1 j_2;j}(\epsilon_c) \ket{\Phi^{ac}_{j_1 j_2 i}}
  +
  \sum_{j_1} \int d\epsilon_c \ d_{j_1 a;j}(\epsilon_c) \ket{\Phi^{c}_{j_1 i}}
  ,
\end{equation}
where the first term is the closed-channel with one Rydberg electron, one outer-shell hole and one inner-shell hole, while the two remaining terms are the spectator (second term) and participator (third term) continuum channels with multiple holes in the outer-shells. 
Additional sub-channels due to different angular momenta are implicitly captured with the integral over all possible continuum states $c$ with energy $\epsilon_c$.
Note that $c$ represents the Auger electron, and $E$ is the energy of the exact eigenstate. 

We assume the final ionic states are stable with a well defined energy. 
A generalization to subsequent Auger decays is possible by heuristically turning the final ionic energy in a Lorentzian distribution centered around the expected final energy and a width given by its decay rate.
To make the expressions more compact, we use $\ket{\Phi_I}=\ket{\Phi^a_{ji}}$ for the closed-channel state, and $\ket{\Phi_F}$ for any open-channel configuration ($\ket{\Phi^{a}_{j_1 j_2 i}}$ or $\ket{\Phi_{j_1 i}}$).
The coefficients read
\begin{eqnarray}
  \label{eq:fano.state.coeff0}
  d_0(E)
  &=&
  \frac{\sin \Delta(E)}{ \sqrt{\pi\Gamma/2} }
  ,
  \\
  \label{eq:fano.state.coeff1}
  d_F(E')
  &=&
  \frac{ V_{I,F}(\epsilon') }{ \sqrt{\Gamma/(2\pi)} }
  \left[
    \frac{1}{\pi} \frac{\sin[\Delta(E)]}{E-E'}
    -
    \cos[\Delta(E)] \delta(E'-E)
  \right]
  ,
\end{eqnarray}
where $\epsilon=E-E_F$ ($\epsilon'=E'-E_F$) is the continuum electron energy with $E_F$ being the energy of the doubly charged ion $\ket{\Phi_F}$, the decay rate is given by $\Gamma=2\pi \sum_F |V_{I,F}(\epsilon)|^2$ with $V_{I,F}(\epsilon)=\bra{I}\hat r^{-1}_{12} \ket{\Phi^\epsilon_F}$, and
\begin{equation}
  \tan[\Delta(E)]
  =
  -\frac{ \Gamma/2 }{ E-E_I-G(E) }
  ,
\end{equation}
where $E_I$ is the energy of close-channel state $\ket{\Phi_I}$ and $G(E)$ is an energy correction due to the coupling between the closed-channel state and the open-channel states (see Ref.~\cite{Fa-PhysRev-1961} for details). 

For an electron far away from the ion with energy $\epsilon$, we know its wavefunction has the form $\braket{{\bf r}}{\varphi_{\Delta}(\epsilon)} \propto k^{-1/2}(\epsilon)\,\sin\big[k(\epsilon)r+\Delta+\delta_\textrm{coul}-0.5\pi l\big]\,Y_{l,m}(\Omega)$, where $l,m$ is the angular momenta of the electron, $k(\epsilon)$ is its asymptotic wave number, and $\delta_\textrm{coul}$ is the Coulomb phase shift.
The asymptotic wavefunction of the continuum electron for the configuration $\ket{\Phi^c_F}$ is given by $\ket{\varphi_{\Delta=0}(\epsilon)}$.
As pointed out in Ref.~\cite{Fa-PhysRev-1961}, the second term in equation~\eqref{eq:fano.state.coeff1} performs the $\Delta$ phase shift when integrated of continuum electron energy in equation~\eqref{eq:fano.state}.
We  write the eigenstate $\ket{E}$ as
\begin{equation}
  \ket{E}
  =
  d_0(E) \ket{\Phi_I}
  +
  \sum_F  \frac{V_{I,F}(\epsilon)}{ \Gamma/(2\pi) }  \ket{\Phi_F;\varphi_{\Delta(E)}(\epsilon)}
  ,
\end{equation}
where $\ket{\Phi_F;c}:=\ket{\Phi^c_F}$.

For an Auger decay, the state $\ket{\Phi_I}$ is the initial state.
The final state, $\ket{\Phi_F^\epsilon}$ , is an outgoing continuum electron in channel $F$ with the asymptotic form $\propto k^{-1/2}(\epsilon)\, e^{i\, k(\epsilon)r}$. 
Since the initial state is not an energy eigenstate, the final Auger electron energy has not one defined energy. 
The transition probability from $\Phi_I$ to a outgoing continuum electron with energy, $\epsilon=E-E_F$, in channel $F$ is given by the overlap between the initial and final configuration,
\begin{eqnarray}
  \beta_{F}(\epsilon)
  &=&
  \braket{\Phi_I}{\Phi_F^\epsilon}
  =
  \int dE' \ \braket{\Phi_I}{E'} \, \braket{E'}{\Phi_F^\epsilon}
  \\
  \nonumber
  &=&
  \int dE'\  d_0(E') \frac{V_{I,F}(E'-E_F)}{ \Gamma/(2\pi) } 
  \underbrace{ \braket{\varphi_{\Delta}(E'-E_F)}{\epsilon} }_{\propto\delta(E'-E_F-\epsilon)}
  \\
  &\propto&
  d_0(E_F+\epsilon) \frac{V_{I,F}(\epsilon)}{ \Gamma/(2\pi) }
  =
  \frac{V_{I,F}(\epsilon)}{ \sqrt{ (\epsilon+E_F-E_I-G)^2 + \Gamma^2/4 } }
  ,
\end{eqnarray}
where we used $\sin(\arctan(x)) = \frac{x}{\sqrt{1+x^2}}$.
The energy correction $G$ is small compared to $E_F$ and $E_I$ so that it can be dropped.
The overlap between the continuum electrons enforces $E'=E_F+\epsilon$ such that both wavefunctions have asymptotically the same wavelength. 
The overlap results also in an additional phase term, which can be ignored as only the probability of the Auger electron is measured.
Replacing $\Phi_I$ and $\Phi_F$ with the original CI-configurations, we get equation~\eqref{eq:state.auger.coeff}.

\section{Auger Spectrum}
\label{app:auger}

The photoelectron spectrum including the Auger decay without making any approximation read
\begin{eqnarray}
  \label{eq:pes.auger}
  \hskip-10ex
  P^\textrm{\scriptsize Auger}(\epsilon,\tau)
  &=&
  \sum_{j,j_1}
  \int\! dc \
  \Big[
  \sum_{j_2} A^\textrm{\scriptsize Auger}_{j;j_1,j_2;c}(\epsilon) 
  + 
  \sum_{p=a,b} A^\textrm{\scriptsize Auger}_{j;j_1 p;c}(\epsilon) 
  +
  2
  B^\textrm{\scriptsize Auger}_{j,j_1,c}(\epsilon)  \cos\Theta^\textrm{\scriptsize Auger}_{j,j_1,c}(\epsilon,\tau)
  \Big]
  \nonumber
  \\
\end{eqnarray}
with
\numparts
\begin{eqnarray}
  \label{eq:pes.auger.coeff1}
  \hskip-10ex
  A^\textrm{\scriptsize Auger}_{j;j_1,p;c}(\epsilon) 
  &=&
  d^2_j(\epsilon)
  \frac{ |V_{j;j_1,p;c}|^2\ |{\cal E}(\epsilon-\epsilon^c_{j_1 p})|^2   }{(\epsilon_c-\epsilon^j_{j_1 p})^2 + \Gamma_j^2/4} 
  ,
  \\
  \label{eq:pes.auger.coeff2}
  \hskip-10ex
  B^\textrm{\scriptsize Auger}_{j,j_1,c}(\epsilon) 
  &=&
  d^2_j(\epsilon)  
  \prod_{p=a,b}
  \Bigg|
  \frac{V_{j;j_1,p;c}\ {\cal E}(\epsilon-\epsilon^c_{j_1 p}) }{\sqrt{(\epsilon_c-\epsilon^j_{j_1 p})^2 + \Gamma_j^2/4}} \
  \Bigg|
  ,
  \\
  \label{eq:pes.auger.coeff3}
  \hskip-10ex
  \Theta^\textrm{\scriptsize Auger}_{j,j_1,c}(\epsilon,\tau) 
  &=&
  \textrm{arg}\Big( \Delta E \tau + [\phi(\epsilon-\epsilon^c_{j_1 b})-\phi(\epsilon-\epsilon^c_{j_1 a})] \Big)
  .
\end{eqnarray}
\endnumparts

Auger resonances are usually not wider than 1--2~eV justifying the approximation for equation~\eqref{eq:pes.auger.simple} that $V_{j;j_1,p;c}$ does not vary across the resonance.
The second assumption was that the phase $\phi(\epsilon-\epsilon^c_{j_1 b})-\phi(\epsilon-\epsilon^c_{j_1 a})$ does also not vary across the resonance. 
This is only true for pulses with chirps, which varies on a tens of eV scale resulting in modification in the pulse duration on the attosecond scale. 
For dispersion relation that lead to femtosecond pulses, this assumption does not hold and the phase variation across the resonance need to be considered as well.

\bibliographystyle{iopart-num}

\bibliography{amo,books,chemistry,solidstate,notes}

\begin{thebibliography}{10}
\bibitem{note-shakeup}
All shake-up amplitudes are calculated for a $3s$ hole making use of the weak
  dependence of the exact shape of the inner hole.
\end{thebibliography}

\end{document}